\documentclass[10pt]{article}
\usepackage[preprint]{tmlr}
\usepackage{hyperref}
\usepackage{url}
\usepackage{graphicx}
\usepackage{booktabs}
\usepackage{multirow}
\usepackage{amsmath,amssymb}
\usepackage{caption}
\usepackage{microtype}
\usepackage{comment}
\graphicspath{{figures/}}
\makeatletter
\if@accepted\newenvironment{nonanon}{}{}\else\excludecomment{nonanon}\fi
\makeatother

\title{Leakage, Recall and the Limits of Temporal Holdouts:\\ Auditing LLM Forecasts of Phase~3 Cancer Trial Completion}

\author{\name Camilla Fezzi \email camillafezzi@gmail.com \\
      \addr Division of Biology and Biological Engineering\\
      California Institute of Technology}

\begin{document}
\maketitle

\begin{abstract}
Large language models (LLMs) are increasingly proposed as forecasters of clinical-trial outcomes. A high
score on past trials, however, can come from outcome information hidden in the input or from outcomes
the model remembers, rather than from reading the design. We audit both on 675 Phase~3 cancer trials from
ClinicalTrials.gov, scoring four LLMs (Claude Opus 5.5, GPT-5.5, Gemini 3.6 Flash and DeepSeek V4.1
Flash) zero-shot through their providers' APIs on identical prompts, and we release every reply.
\emph{(1) The registry leaks the outcome.} The enrolment recorded for a stopped trial is the number of
patients enrolled before it stopped. Using registry version histories, we find that the ratio of actual to
planned enrolment alone ranks trials at an AUC of $0.755$, above a gradient-boosted tree on all fields
($0.716$), while planned enrolment carries no signal ($0.455$).
\emph{(2) The models know the trials.} From design fields alone they return the exact NCT identifier for
43--69\% of trials, including 36--66\% of those whose results were posted from 2025 on, and a prompt
containing only the identifier and title comes within $0.016$ AUC of the full record without enrolment. A
holdout stratified by when results were posted therefore cannot test memorisation, and a non-LLM control
cannot rescue it, because the leak itself changes across strata.
\emph{(3) Skill survives, but smaller.} On 136 trials whose status first became final after GPT-5.5's
December 2025 cutoff, AUC falls to $0.57$--$0.72$ (Claude $0.722$ [$0.636$, $0.804$]; GPT-5.5 $0.647$).
Claude's lead over GPT-5.5 is smallest ($0.01$) on trials only Claude could have seen and largest ($0.24$)
on trials neither could have seen, the opposite of what memorisation predicts, and two comparisons with
Gemini, whose cutoff falls in between, point the same way less precisely. Probability quality varies
widely: DeepSeek answers 0 or 100 for 85\% of trials and scores worse than the base rate. We recommend
auditing every input against registry histories, probing recognition directly, and reporting skill on
outcomes dated after each model's training cutoff.
\end{abstract}

\section{Introduction}
\label{sec:intro}

Can a large language model forecast whether a clinical trial will run to completion? LLMs already read
trial protocols and match patients to trials \citep{jin2024trialgpt}, and they are increasingly proposed
as tools for testing a protocol before it runs. A model that could flag weak designs in advance would save
years of work and large sums. A high score on a retrospective test set, however, can arise in at least
three ways that have nothing to do with reading a design:

\begin{enumerate}\itemsep2pt
  \item \textbf{Input leakage.} A field supplied as a ``design'' feature may be recorded after the
  outcome is known.
  \item \textbf{Recall.} The model may have seen the trial and its outcome during training.
  \item \textbf{Metric choice.} When roughly three in four trials complete, a model can rank trials well
  and still issue probabilities worse than a constant.
\end{enumerate}

Trial registries are a good setting in which to separate these explanations. Outcomes are objective and
dated, and ClinicalTrials.gov keeps every posted version of every record, so each field can be inspected
as it stood at registration, before any outcome existed.

A common answer to the recall problem is a temporal holdout: stratify the test trials by when their
results were posted, check that skill holds up in the newest stratum, and use a model that cannot
memorise as a control for difficulty. On our data this design appears to clear the LLMs of memorisation.
We show why it cannot. Every test trial reached its primary completion before Claude's training cutoff,
and all but one before GPT-5.5's. Every model names most trials by their registry identifier, including
those in the newest stratum. And the leak itself shifts across strata, so the control declines for
mechanical reasons. We then score the models on trials resolved after their cutoffs and compare models
whose cutoffs differ.

\paragraph{Contributions.}
\begin{itemize}\itemsep2pt
  \item \textbf{A measured leak.} Using registry version histories, we show that actual enrolment, a
  field routinely treated as design information, carries most of a tree model's skill, and that the
  ratio of actual to planned enrolment alone outranks the tree (Section~\ref{sec:leak}).
  \item \textbf{Direct evidence of recall.} A recognition probe and a prompt ablation, run on four model
  families, show that the models identify most trials and recall many outcomes
  (Sections~\ref{sec:recog} and~\ref{sec:ablation}).
  \item \textbf{Why results-posting holdouts fail}, with or without a non-LLM control
  (Section~\ref{sec:strata}).
  \item \textbf{A post-cutoff evaluation and cross-model window tests} that compare models with
  different published training cutoffs on the same unseen trials (Sections~\ref{sec:post}
  and~\ref{sec:window}).
  \item \textbf{A replicated, fully logged evaluation}: four models queried through first-party APIs and
  checked against an earlier independent collection run, with seven prompt variants, run-to-run repeats,
  calibration and label sensitivity, and every prompt, reply and script released.
\end{itemize}

\section{Related Work}
\label{sec:related}

\paragraph{LLMs as surrogates for human judgement.}
A fast-growing literature asks whether LLMs can stand in for human respondents and decision-makers.
\citet{argyle2023out} showed that language models conditioned on demographic backstories reproduce survey
response distributions; \citet{park2024generative} built generative agents whose behaviour tracked a
thousand real interviewees; and \citet{manning2025general} built ``general social agents'' that predict
human play in novel games better than standard behavioural models. Against this optimism,
\citet{gao2025caution} urged caution about treating LLMs as human surrogates, and \citet{bail2024can}
asked whether generative AI can improve social science at all. We inherit this framing but change the
question: rather than asking whether an LLM mimics a human, we ask whether its prediction about a public
event rests on the information it was given.

\paragraph{Machine learning for clinical trial outcome prediction.}
Predicting trial outcomes is an established task. \citet{fu2022hint} introduced HINT and the TOP
benchmark, and \citet{chen2025trialbench} released TrialBench, 23 datasets spanning eight trial
prediction tasks. LLM-specific work is newer and more mixed. \citet{reinisch2024ctp} proposed CTP-LLM, a
fine-tuned GPT-3.5 model that reads protocol text; LLM4TOP \citep{qian2025llm4top} fine-tuned Qwen3-8B
with LoRA and recast outcome prediction as binary classification; and in a related clinical task,
\citet{tan2026ml} found classical machine learning more accurate than LLMs for structured survival
prediction. This literature reports discrimination almost exclusively. HINT and TrialBench report AUROC,
PRAUC and F1, and LLM4TOP notes that ``a higher AUROC does not inevitably imply better-calibrated
probability predictions'' \citep{qian2025llm4top}. We know of no trial-outcome study that reports a Brier
score or a reliability curve; the closest work on uncertainty \citep{lu2024uncertainty} evaluates with
ranking metrics and F1, none of which measures calibration. Nor do we know of one that audits its inputs
against registry histories or tests whether the model recognises the trials.

\paragraph{Calibration and confidence in LLMs.}
Instruction-tuned LLMs are often poorly calibrated. \citet{leng2025reward} trace this to reward models
that favour confident responses, \citet{xiao2025restoring} to preference collapse during alignment, and
\citet{sanzguerrero2026overconfident} separate the effects of instruction tuning from those of the chat
template. Verbalised confidence can outperform token probabilities for instruction-tuned models
\citep{tian2023just} but is itself prone to overconfidence \citep{xiong2024can}, and
\citet{wang2024calibrating} show that rescaling verbalised probabilities naively distorts them. In
clinical prediction, calibration is treated as a requirement rather than an option
\citep{huang2020tutorial, dehond2022guidelines}.

\paragraph{Contamination and temporal holdouts.}
Worse performance on material released after a model's training cutoff is widely read as evidence of
contamination, both in coding \citep{jain2024livecodebench} and in dated benchmarks generally.
\citet{zhang2026testoftime} show that this temporal signal depends on how benchmark items are
constructed, even when the source material is held fixed. Our setting adds a further problem: the date
used to split the data (when results were posted) is not the date the outcome became knowable, and the
split also changes what a leaking input reveals. We therefore date outcomes against each model's cutoff
and compare models whose cutoffs differ.

\section{Data}
\label{sec:data}

\subsection{Source, sample and labels}
We queried the ClinicalTrials.gov API v2 on \textbf{8 October 2026} for all studies matching
\texttt{AREA[Phase]PHASE3 AND AREA[ConditionSearch]cancer AND AREA[HasResults]true}. Restricting the
sample to studies with posted results guarantees known outcomes but makes the sample selected
(Section~\ref{sec:limitations}). The query returned 2{,}603 studies. Keeping those that are completed,
terminated or withdrawn and have a usable start year leaves \textbf{2{,}366 trials: 1{,}932 completed and
434 stopped}, all of them terminated. For each trial we extracted its identifiers and title; its status
and free-text reason for stopping; phase, enrolment, arms, allocation and purpose; sponsor class;
conditions; interventions; primary outcome measures; key dates; and eligibility criteria.

The primary label is completed versus stopped. A stopped trial has not necessarily failed: the reason
for stopping, recorded for 391 of the 434 stops, cites efficacy or futility in only 15.7\% of them
(Table~\ref{tab:audit}). The label therefore measures \emph{non-completion}, not drug failure. We also
score two stricter definitions (Table~\ref{tab:labels}).

\begin{table}[t]
\centering
\caption{Reported reasons for stopping ($n=434$ terminated trials), keyword-classified in the priority order shown. ``Lack of'' counts as efficacy only with an efficacy word (``lack of efficacy'', ``benefit'', ``effect'', ``activity'' or ``response'').}
\label{tab:audit}
\small
\begin{tabular}{lrr}
\toprule
Reported reason (priority order) & Trials & Share of stops \\
\midrule
Efficacy or futility & 68 & 15.7\% \\
Safety or toxicity & 36 & 8.3\% \\
Accrual or enrolment problems & 122 & 28.1\% \\
Administrative / business / funding & 72 & 16.6\% \\
Other / not classifiable by keyword & 93 & 21.4\% \\
No reason recorded & 43 & 9.9\% \\
\bottomrule
\end{tabular}
\end{table}

\begin{table}[t]
\centering
\caption{Label definitions and base rates. Test = trials starting after 2015. B drops any stop whose reason mentions accrual, enrolment, recruitment, administration, business, funding, sponsor, strategy or finance, so it removes more stops than the accrual and administrative rows of Table~\ref{tab:audit}.}
\label{tab:labels}
\small
\begin{tabular}{clrrrr}
\toprule
Def & Definition & $N$ & Base rate & Test $n$ & Test base rate \\
\midrule
\textbf{A} & Completed vs.\ all stopped \emph{(primary)} & 2{,}366 & 0.817 & 675 & 0.754 \\
\textbf{B} & Drop stops that mention accrual, administration or sponsor & 2{,}139 & 0.903 & 578 & 0.881 \\
\textbf{C} & Completed vs.\ explicit efficacy/futility stops only & 2{,}000 & 0.966 & 541 & 0.941 \\
\bottomrule
\end{tabular}
\end{table}

\subsection{Splits and the results-posting strata}
The tree model trains on the trials that started by 2015 ($n=1{,}691$, completion rate $0.842$), and every
forecaster is evaluated on the \textbf{675 trials that started later} (166 stopped, completion rate
$0.754$). A common proxy for when an outcome could have entered a model's training data is the year its
results were first posted. We stratify the test trials this way, into results posted by 2022 ($n=295$),
in 2023--24 ($n=243$) and from 2025 on ($n=137$), in order to show why such strata fail as a memorisation
test (Section~\ref{sec:strata}).

\subsection{Registry histories}
On 11 October 2026 we retrieved the full version history of all 2{,}366 trials, and of the post-cutoff
candidates described below, from ClinicalTrials.gov's record history, which serves each posted version of
a record with its date; we release the snapshot. From it we built two further views of each trial: the
record \emph{as first registered} (version 0), which carries planned rather than actual enrolment, and
the last version \emph{before the outcome}, which precedes the first change of status to completed or
terminated. Comparing these versions field by field with the current record shows which inputs can
change after the outcome.

\subsection{Trials resolved after the training cutoffs}
\label{sec:postdata}
Every test trial reached its primary completion by 16 December 2025, before Claude's training cutoff
and, with one exception, before GPT-5.5's. The registry's status field lags behind this date. Sixty-five
test trials (10, 20 and 35 in the three strata) first showed a final status on or after 1 December 2025,
and 11, all of them completed, on or after 1 July 2026; 46 of the 65 had posted results before
1 December 2025. These trials remain in the test set, which we treat as in-sample throughout.

To obtain unseen outcomes, we dropped the results filter and took the Phase~3 trials from the same
condition search that were updated after 1 December 2025. Of these, 243 first showed a completed or
terminated status on or after 1 December 2025 (51 terminated), and 65 on or after 1 July 2026 (10
terminated). We kept the 218 with a cancer term in their conditions or title and dropped the 82 already in
our 2{,}366-trial dataset (62 of them test trials, 54 of which had posted results before their status
changed). This leaves \textbf{136 trials (35 stopped)} that became final from 1 December 2025 and
\textbf{43 (9 stopped)} that became final from 1 July 2026. Each is prompted with its last record version
before the status change, with enrolment withheld. A stricter version of each set also requires the
trial's own primary completion to fall no earlier than six months before the cutoff.

\section{Methods}
\label{sec:methods}

Every forecaster is scored on the same trials with the same labels. Intervals are 95\% percentile
bootstrap intervals with 2{,}000 resamples, and differences between forecasters are paired, computed on
the same resample.

\paragraph{Baselines.}
The \emph{base rate} assigns every trial the training completion rate ($0.842$). The
\emph{gradient-boosted tree} is an XGBoost model (300 trees, depth 4, learning rate 0.05, subsample and
column subsample 0.8, seed 0) trained on the pre-2016 split, using log enrolment, arm count, start year,
intervention count, eligibility length, an overall-survival endpoint indicator, sponsor class,
intervention types and the 25 most common conditions. We also fit the tree with \emph{planned} enrolment
from the first registered version, and a TF-IDF text model (word and character $n$-grams with logistic
regression) on the prompt text. The text model is trained on the pre-2016 split or cross-fitted on the
test trials, either with random folds or with folds that keep each drug programme on one side. A rolling
origin trains on trials that started by 2018 and tests on the 286 test trials that started in 2019 or
later.

\paragraph{LLM forecasters.}
We queried the four models through their providers' own APIs between 12 and 16 October 2026
(Table~\ref{tab:models}). Each prompt supplies phase, conditions, interventions, primary endpoints,
enrolment, sponsor class, arm count, allocation, primary purpose and up to 6{,}000 characters of
eligibility criteria. It omits the title and NCT identifier and ends with a required answer line,
\texttt{ANSWER: <integer from 0 to 100>} (Appendix~\ref{app:prompt}), which is the only part of the
reply the parser reads. Every reply is stored in full, together with any reasoning text, the finish
reason, token counts, the model version reported by the API and the request identifiers. Failed or empty
calls were retried. Six of DeepSeek's 7{,}068 calls (three recognition probes and three eligibility-only
prompts) never returned an answer because the model's reasoning exceeded the 32{,}000-token cap.

\begin{table}[t]
\centering
\caption{Models, settings and usage (first-party APIs, 12--16 October 2026). Training cutoffs as published by each provider \citep{anthropic2026opus, openai2026gpt55, google2026gemini36}. Calls: answered calls out of 7{,}068 per model. Tokens: input / output in millions; output includes reasoning tokens.}
\label{tab:models}
\small
\setlength{\tabcolsep}{3.5pt}
\begin{tabular}{llp{1.9cm}p{4.0cm}rr}
\toprule
Model & Version that answered & Training cutoff & Settings & Calls & Tokens (M) \\
\midrule
Claude Opus 5.5 & \texttt{claude-opus-5-5} & Jun 2026 & Temperature not settable (API default); 2{,}048-token cap & 7{,}068 & 7.6 / 1.6 \\
GPT-5.5 & \texttt{gpt-5.5-2026-04-23} & 1 Dec 2025 & Default reasoning effort; 8{,}000-token cap & 7{,}068 & 4.5 / 7.0 \\
Gemini 3.6 Flash & \texttt{gemini-3.6-flash}$^{a}$ & Mar 2026$^{c}$ & Temperature 0; 4{,}096-token cap & 7{,}068 & 4.5 / 5.9 \\
DeepSeek V4.1 Flash & \texttt{deepseek-flash}$^{b}$ & not stated & Thinking mode (default); temperature 0 sent; 32{,}000-token cap & 7{,}062 & 4.5 / 32.6 \\
\bottomrule
\end{tabular}

\vspace{2pt}\raggedright\footnotesize $^{a}$Requested as \texttt{gemini-3.5-flash}; Google now serves 3.6 Flash for that identifier. $^{b}$DeepSeek V4.1 Flash on DeepSeek's own API. $^{c}$The model card (July 2026) notes that some domains may be limited to January 2025. Six GPT-5.5 and eight DeepSeek answers come from test calls with lower caps (2{,}048 and 8{,}000 tokens); all finished normally.
\end{table}

\paragraph{Prompt variants and the recognition probe.}
All four models answer seven variants on the 675 test trials: the full current record (\textbf{C}); the
same without enrolment (\textbf{N}); structured fields only, with enrolment but without eligibility text
(\textbf{A}); eligibility text only (\textbf{B}); condition and interventions only (\textbf{X}); the NCT
identifier and title only, with an invitation to use what the model remembers (\textbf{I}); and a
\textbf{recognition probe} (\textbf{R}). The probe gives the design fields without identifier, title or
enrolment and asks, in JSON, whether the model recognises the trial and, if so, its name, NCT identifier
and outcome. Variants C and N are also run on the record as first registered, and N on the post-cutoff
trials. Five extra runs of variant C on a stratified random subset of 150 test trials measure run-to-run
variation.

\paragraph{Window tests.}
The providers publish training cutoffs of 1 December 2025 for GPT-5.5, March 2026 for Gemini 3.6 Flash
and June 2026 for Claude Opus 5.5; DeepSeek does not state one (Table~\ref{tab:models}). A trial whose
status first became final between two cutoffs could have been seen by the later model but not by the
earlier one. If remembered outcomes drove a model's advantage, its lead should be larger on such trials
than on trials that became final from 1 July 2026, which no model could have seen. We compare Claude with
GPT-5.5 on trials that became final between 1 December 2025 and 30 June 2026 (93 trials, 26 stopped),
Claude with Gemini on those that became final between 1 April and 30 June 2026 (43 trials, 11 stopped),
and Gemini with GPT-5.5 on those that became final between 1 December 2025 and 31 March 2026 (50 trials,
15 stopped). Each window is set against the 43 trials (9 stopped) that became final from 1 July 2026. The
statistic is the lead in the window minus the lead after it, with an interval from 4{,}000 independent
resamples of the two sets.

\paragraph{Calibration and metrics.}
We report the area under the ROC curve (AUC), the Brier score with its base-rate reference, the expected
calibration error (ECE, 10 equal-width bins) and the mean forecast. Platt scaling, a logistic regression
on the logit of the forecast clipped to $[0.005, 0.995]$, is fitted by two-fold cross-fitting, so every
calibrated forecast is out of sample.

\section{Results}
\label{sec:results}

\subsection{The enrolment field records the outcome}
\label{sec:leak}

\begin{figure}[t]
\centering
\includegraphics[width=\textwidth]{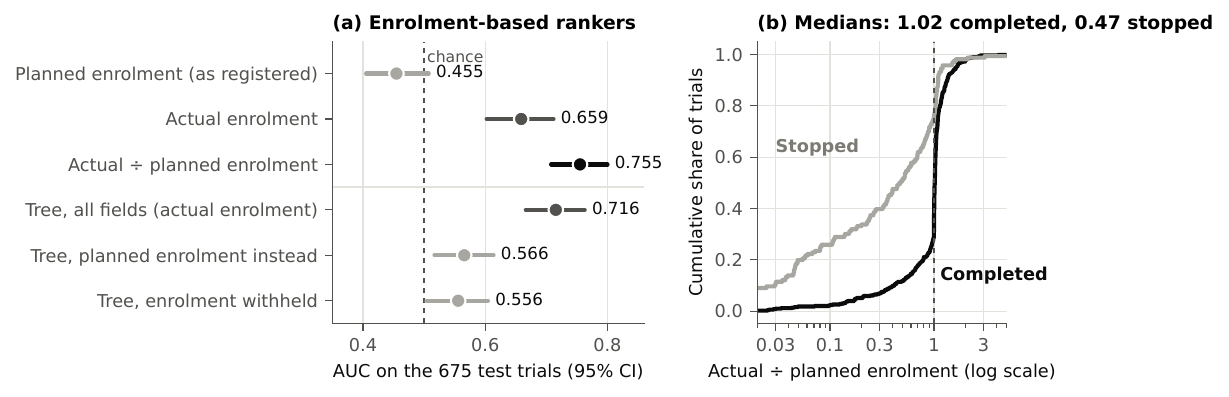}
\caption{(a) AUC of enrolment-based rankers and of the gradient-boosted tree on the 675 test trials,
with 95\% bootstrap intervals; planned enrolment is taken from each trial's first registered version.
(b) Cumulative distribution of actual $\div$ planned enrolment by outcome.}
\label{fig:leak}
\end{figure}

For a completed or terminated trial, the registry reports \emph{actual} enrolment, which for a stopped
trial is the number of patients enrolled before it stopped. Median actual enrolment is 296 in completed
test trials and 96 in stopped ones. The registry histories let us measure this leak directly
(Figure~\ref{fig:leak}). On the 674 test trials with a planned figure, actual enrolment ranks the trials
at an AUC of $0.659$ [$0.602$, $0.712$], and the ratio of actual to planned enrolment at $0.755$ [$0.708$,
$0.800$]. Planned enrolment, the figure known at registration, scores $0.455$ [$0.405$, $0.507$]. The median
ratio is $1.02$ for completed trials and $0.47$ for stopped ones.

The tree is almost entirely an enrolment model. Log enrolment dominates its permutation importance
($0.207$ AUC, roughly ten times the next feature), its predictions correlate with log enrolment at
$r=0.83$, and withholding the field costs it $0.160$ [$0.104$, $0.216$] AUC, from $0.716$ to $0.556$.
Replacing actual with planned enrolment does not restore it ($0.566$). Text models on the prompt do little
better ($0.580$--$0.597$) unless they are cross-fitted with random folds on the test trials themselves
($0.667$), which lets trials from the same drug programme inform each other (Table~\ref{tab:ladder}).
The LLMs lose much less from withholding enrolment: $0.024$ (Claude), $0.030$ (GPT-5.5), $0.040$
(Gemini) and $0.051$ (DeepSeek), with every interval excluding zero. Apart from enrolment, the edits made
after the outcome carry little information for the models. Without enrolment, AUC on the current record
minus AUC on the record as first registered ranges from $-0.014$ to $+0.010$, and no interval excludes
zero (Table~\ref{tab:replication}). With enrolment, the difference is $+0.035$ to $+0.045$; because the
first registered record shows planned enrolment, this is the same leak seen from the other side.

\subsection{Four models on the same trials}
\label{sec:h2h}

\begin{table}[t]
\centering
\caption{Every forecaster on the same 675 test trials (completion rate 0.754; a constant forecast at that rate has a Brier score of 0.185). AUC with 95\% bootstrap intervals (2{,}000 resamples). Platt: two-fold cross-fitted Platt scaling. Full record: the current registry record. Enrolment withheld: the same prompt without the enrolment line; the tree is retrained without the field.}
\label{tab:h2h}
\small
\setlength{\tabcolsep}{4pt}
\resizebox{\textwidth}{!}{%
\begin{tabular}{lcccccccc}
\toprule
 & \multicolumn{4}{c}{Full record} & \multicolumn{3}{c}{Enrolment withheld} \\
\cmidrule(lr){2-5}\cmidrule(lr){6-8}
Forecaster & AUC & Brier & ECE & Mean $p$ & AUC & Brier & ECE \\
\midrule
Base rate (training, 0.842) & 0.500 & 0.193 & 0.087 & 0.842 & & & \\
Gradient-boosted tree & 0.716 [0.666, 0.763] & 0.159 & 0.084 & 0.807 & 0.556 [0.504, 0.604] & 0.197 & 0.106 \\
\quad Platt & 0.715 & 0.155 & 0.013 & 0.756 & & & \\
\midrule
Claude Opus 5.5 & 0.923 [0.900, 0.943] & 0.101 & 0.096 & 0.658 & 0.899 [0.872, 0.924] & 0.113 & 0.097 \\
\quad Platt & 0.921 & 0.091 & 0.026 & 0.755 & 0.898 & 0.106 & 0.031 \\
GPT-5.5 & 0.874 [0.843, 0.901] & 0.129 & 0.064 & 0.714 & 0.844 [0.812, 0.876] & 0.142 & 0.061 \\
\quad Platt & 0.872 & 0.123 & 0.053 & 0.754 & 0.840 & 0.137 & 0.072 \\
Gemini 3.6 Flash & 0.868 [0.837, 0.897] & 0.130 & 0.083 & 0.694 & 0.828 [0.796, 0.862] & 0.150 & 0.080 \\
\quad Platt & 0.856 & 0.125 & 0.039 & 0.754 & 0.798 & 0.146 & 0.044 \\
DeepSeek V4.1 Flash & 0.767 [0.728, 0.803] & 0.223 & 0.219 & 0.673 & 0.716 [0.672, 0.759] & 0.258 & 0.251 \\
\quad Platt & 0.755 & 0.148 & 0.024 & 0.754 & 0.705 & 0.162 & 0.013 \\
\bottomrule
\end{tabular}}
\end{table}

On the full record, Claude Opus 5.5 ranks the test trials at an AUC of $0.923$, GPT-5.5 at $0.874$, Gemini
at $0.868$ and DeepSeek at $0.767$ (Table~\ref{tab:h2h}). Paired against the tree, the first three are
clearly ahead (Claude $+0.208$ [$+0.160$, $+0.258$]; GPT-5.5 $+0.159$; Gemini $+0.152$), but DeepSeek is
not ($+0.052$ [$-0.009$, $+0.112$]). Without enrolment, Claude leads GPT-5.5 by $0.055$ [$0.032$, $0.080$],
Gemini by $0.070$ and DeepSeek by $0.182$, while GPT-5.5 and Gemini are indistinguishable. Averaging the
four forecasts does not help: the mean ($0.860$) is worse than Claude alone by $0.039$ [$0.023$, $0.057$].
The rolling origin gives the same picture. On the 286 trials that started from 2019, the tree without
enrolment scores $0.606$ and Claude $0.883$ (Table~\ref{tab:ladder}).

These results replicate. An earlier, independent collection run on 8 October 2026, made through a
third-party API router with a looser answer parser, gives nearly the same AUCs for Claude ($0.921$
against $0.923$ here; per-trial $r=0.96$), GPT-5.5 ($0.868$ against $0.874$; $r=0.89$) and DeepSeek ($0.749$
against $0.767$; $r=0.67$). Gemini's identifier was served by a newer model in this run, and its AUC rose
from $0.815$ to $0.868$ (Table~\ref{tab:replication}).

All of these numbers are in-sample. Every test trial reached primary completion before Claude's training
cutoff and all but one before GPT-5.5's, and by the status field all but 11 had become final before
Claude's cutoff (Section~\ref{sec:postdata}). The next three sections ask how much of the skill is recall.

\subsection{The models recognise the trials}
\label{sec:recog}

\begin{table}[t]
\centering
\caption{The recognition probe, which shows design fields only, with the NCT number, title and enrolment removed (Appendix~\ref{app:prompt}). Says yes: the model reports recognising the trial. Exact NCT: it returns the trial's own eight-digit identifier, also split by results-posting stratum (295 / 243 / 137 trials). Recall right: share of definite recalled outcomes that match the registry. Recall AUC: the recalled outcome used as a forecast (0.5 when not recognised). Last two columns: AUC without enrolment on the trials the model names exactly and on the rest. DeepSeek answered 672 of the 675 probe calls (three hit the token cap).}
\label{tab:recog}
\small
\setlength{\tabcolsep}{3.5pt}
\resizebox{\textwidth}{!}{%
\begin{tabular}{lrcccccc}
\toprule
Model & Says yes & Exact NCT & by stratum & Recall right & Recall AUC & AUC, named & AUC, not named \\
\midrule
Claude Opus 5.5 & 604 & 468 (69\%) & 200 / 178 / 90 & 87\% & 0.818 [0.782, 0.853] & 0.919 & 0.849 \\
GPT-5.5 & 616 & 426 (63\%) & 183 / 163 / 80 & 79\% & 0.805 [0.768, 0.838] & 0.860 & 0.819 \\
Gemini 3.6 Flash & 615 & 289 (43\%) & 118 / 118 / 53 & 82\% & 0.774 [0.738, 0.812] & 0.841 & 0.817 \\
DeepSeek V4.1 Flash & 671 & 313 (46\%) & 138 / 125 / 50 & 72\% & 0.720 [0.679, 0.759] & 0.712 & 0.716 \\
\bottomrule
\end{tabular}}
\end{table}

\begin{figure}[t]
\centering
\includegraphics[width=\textwidth]{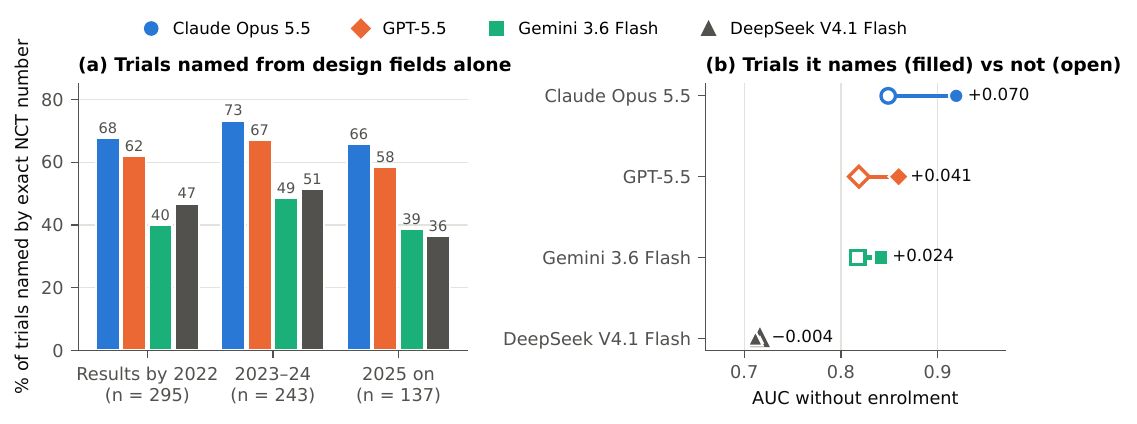}
\caption{(a) Share of trials for which each model returns the exact NCT identifier from design fields
alone, by results-posting stratum. (b) AUC without enrolment on trials the model names (filled) and on
the rest (open); the label gives the difference.}
\label{fig:recog}
\end{figure}

Given only the design fields, with identifier, title and enrolment removed, Claude returns the trial's
exact eight-digit NCT identifier for 468 of the 675 trials (69\%), GPT-5.5 for 426 (63\%), DeepSeek for
313 (46\%) and Gemini for 289 (43\%) (Table~\ref{tab:recog}, Figure~\ref{fig:recog}a). An identifier of
this form cannot be guessed from a design, so these trials are recalled. The share does not fall in the
stratum whose results were posted from 2025 on: Claude names 66\% of it. Self-reported recognition, by
contrast, is uninformative, since every model claims to recognise 89--99\% of trials; we therefore rely on
the exact identifier.

The models also recall outcomes. Used as a forecast, the outcome each model says it remembers ranks the
trials at an AUC of $0.72$--$0.82$ and matches the registry in 72--87\% of definite answers. Every model
over-recalls stopping: of the trials a model remembers as terminated, between 30\% (Claude) and 53\%
(DeepSeek) in fact completed. Recall feeds the forecasts unevenly (Figure~\ref{fig:recog}b). Without
enrolment, Claude ranks the trials it names at $0.919$ and the rest at $0.849$, whereas for DeepSeek the
two are the same ($0.712$ and $0.716$). Naming is only a lower bound on recall, so even the 178 trials that
no model names (45 stopped) are not clean. On them, without enrolment, the models score $0.819$ (Claude),
$0.768$ (GPT-5.5), $0.761$ (Gemini) and $0.657$ (DeepSeek).

\subsection{Where the signal comes from: the prompt ablation}
\label{sec:ablation}

\begin{figure}[t]
\centering
\includegraphics[width=\textwidth]{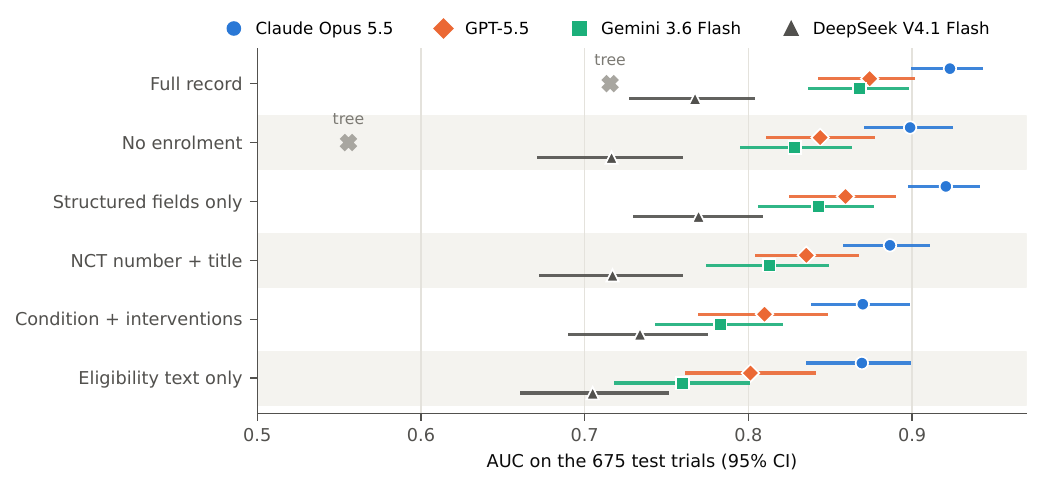}
\caption{AUC by prompt variant for each model on the 675 test trials, with 95\% bootstrap intervals.
Crosses mark the gradient-boosted tree with and without enrolment. Values in Table~\ref{tab:ablation}.}
\label{fig:ablation}
\end{figure}

The prompt ablation points the same way (Figure~\ref{fig:ablation}, Table~\ref{tab:ablation}). Given only
the NCT identifier and title, with an invitation to use what it remembers, every model comes within
$0.016$ of its no-enrolment AUC: Claude scores $0.887$ against $0.899$, GPT-5.5 $0.835$ against $0.844$,
Gemini $0.813$ against $0.828$ and DeepSeek $0.717$ against $0.716$, and no paired difference excludes
zero. Once a model has the trial's name, the rest of the design (eligibility, endpoints, sponsor, arms
and allocation) adds nothing measurable. Structured fields alone come within $0.026$ of the full record
for every model, so the eligibility text also adds little. The condition and interventions alone reach
$0.73$--$0.87$, a figure that mixes legitimate knowledge about drug classes with knowledge of the specific
trial. Because the design alone cannot separate the two, we turn to outcomes the models cannot have seen.

\subsection{Why the results-posting holdout cannot detect memorisation}
\label{sec:strata}

\begin{figure}[t]
\centering
\includegraphics[width=\textwidth]{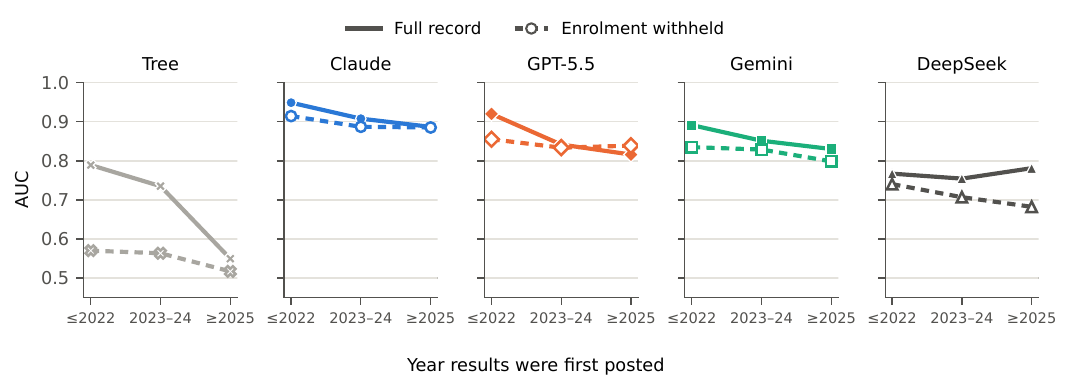}
\caption{AUC by the year results were first posted, with the full record (solid) and with enrolment
withheld (dashed). The tree's decline follows the enrolment field; the LLMs' curves without enrolment
are much flatter. Values in Table~\ref{tab:strata}.}
\label{fig:strata}
\end{figure}

Read naively, Figure~\ref{fig:strata} is evidence against memorisation. AUC falls for every forecaster
from the oldest to the newest stratum, but the tree, which cannot memorise anything, falls the most
($0.789 \to 0.550$). Three facts undo this reading.

\emph{First, the newest stratum is not unseen.} The date results were posted is not the date of the
outcome. Every test trial reached its primary completion by 16 December 2025, before Claude's training
cutoff, and only 4 of the 137 newest-stratum trials first showed a final status after it. The models also
name the newest stratum's trials as often as the oldest (Figure~\ref{fig:recog}a).

\emph{Second, the tree's decline is mechanical.} Results are posted a median of 12 months after
completion in every stratum, so the strata differ mainly in which stopped trials they contain. Stopped
trials with results posted from 2025 on ran longer than those posted by 2022 (median 38 against 22 months)
and enrolled more patients (358 against 63). As a result, enrolment stops giving the outcome away. Median
actual enrolment is 311 for completed trials against 63 for stopped ones in the oldest stratum, but 228.5
against 358 in the newest, where enrolment alone ranks the trials at chance (AUC $0.490$, against $0.739$
in the oldest). The ratio of actual to planned enrolment still separates the newest stratum's trials
($0.642$), but neither the tree nor the models see planned enrolment. The tree tracks the field, not
general difficulty.

\emph{Third, the LLMs' curves are flat for the wrong reason.} Without enrolment, Claude's AUC is $0.915$,
$0.887$ and $0.886$ across the three strata. A flat curve rules out memorisation only if some stratum lies
after the training cutoff. Here none does, and recall in every stratum would produce the same picture.
The gain from enrolment, by contrast, is concentrated where the leak is strongest: in the oldest stratum it
adds $0.034$ (Claude), $0.065$ (GPT-5.5) and $0.056$ (Gemini), each excluding zero, while in the newest
it adds nothing distinguishable from zero for those three models.

\subsection{Trials resolved after the training cutoffs}
\label{sec:post}

\begin{table}[t]
\centering
\caption{AUC on trials whose registry status first became completed or terminated after a given date, each prompted with its last record version before that change and without enrolment. Cancer conditions only, excluding trials already in the 2{,}366-trial dataset. 1 December 2025 is GPT-5.5's cutoff; 1 July 2026 follows Claude's. The tree is trained on trials that started by 2015, without enrolment. 95\% bootstrap intervals.}
\label{tab:post}
\small
\setlength{\tabcolsep}{4pt}
\begin{tabular}{lcccc}
\toprule
 & Test set & From 1 Dec 2025 & From 1 Jul 2026 & 1 Dec 2025 -- 30 Jun 2026 \\
Model & (675; 166 stopped) & (136; 35 stopped) & (43; 9 stopped) & (93; 26 stopped) \\
\midrule
Claude Opus 5.5 & 0.899 & 0.722 [0.636, 0.804] & 0.824 [0.688, 0.935] & 0.679 [0.563, 0.783] \\
GPT-5.5 & 0.844 & 0.647 [0.546, 0.743] & 0.587 [0.346, 0.792] & 0.668 [0.554, 0.778] \\
Gemini 3.6 Flash & 0.828 & 0.590 [0.488, 0.687] & 0.521 [0.304, 0.730] & 0.622 [0.504, 0.731] \\
DeepSeek V4.1 Flash & 0.716 & 0.574 [0.479, 0.671] & 0.634 [0.437, 0.802] & 0.547 [0.428, 0.660] \\
\midrule
Gradient-boosted tree & 0.556 & 0.566 [0.455, 0.672] & 0.507 [0.254, 0.756] & 0.560 [0.425, 0.690] \\
\bottomrule
\end{tabular}
\end{table}

\begin{figure}[t]
\centering
\includegraphics[width=\textwidth]{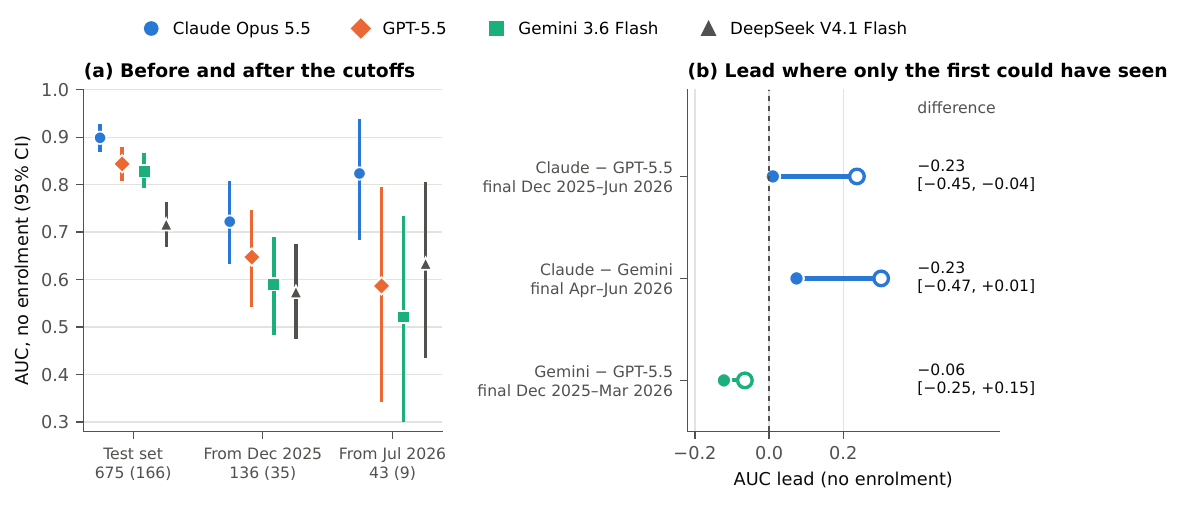}
\caption{(a) AUC without enrolment on the test set and on trials whose status first became final from
1 December 2025 or from 1 July 2026; $n$ (stopped) under each group. (b) Window tests. Each row shows the
AUC lead of the first model over the second on trials resolved in a window that only the first could
have seen (filled) and on the 43 trials that became final from 1 July 2026, which neither could have seen
(open). Memorisation predicts filled to the right of open. Right: the difference, window minus after, with
its 95\% interval (Table~\ref{tab:windows}).}
\label{fig:post}
\end{figure}

On trials whose status first became final after GPT-5.5's cutoff, every model loses ground
(Table~\ref{tab:post}, Figure~\ref{fig:post}a). Relative to its test-set AUC without enrolment, Claude
falls by $0.177$ to $0.722$ [$0.636$, $0.804$], GPT-5.5 by $0.197$ to $0.647$ [$0.546$, $0.743$], Gemini by
$0.238$ to $0.590$ and DeepSeek by $0.142$ to $0.574$. Claude and GPT-5.5 stay above chance, while the
intervals for Gemini and DeepSeek include $0.5$. On the 43 trials that became final from July 2026, after
Claude's cutoff, Claude reaches $0.824$ [$0.688$, $0.935$], but this rests on 9 stopped trials; the other
models fall to $0.52$--$0.63$. The tree without enrolment scores $0.566$ [$0.455$, $0.672$] on the 136
trials, or $0.594$ when trained on all 2{,}366 registered trials. Claude's lead over it, $+0.156$
[$-0.007$, $+0.305$], is large, but its interval just includes zero. A stricter set, which also requires
each trial's primary completion to fall no earlier than six months before the cutoff, gives the same
ordering of models (Table~\ref{tab:strict}).

The drop admits two readings: lost memory, or a harder set of trials. These trials were still running in
late 2025, are judged on their pre-outcome records, and differ in kind from the main sample. Twenty-seven
of the 136 reached primary completion before 2024, and of the 9 stopped after July 2026, only one cites
an interim analysis; the rest cite sponsor decisions, funding, accrual or drug supply. The window tests
separate the two readings.

\subsection{Does memory explain the differences between models? The window tests}
\label{sec:window}

Claude falls about as far as GPT-5.5 on the December set, even though two-thirds of those trials (93 of
136) resolved before Claude's own cutoff. On those 93 trials, Claude's lead over GPT-5.5 is $+0.010$
[$-0.068$, $+0.092$]; on the 43 trials neither could have seen, it is $+0.237$ [$+0.064$, $+0.437$]; and on
the test set, $+0.055$. Memorisation predicts the largest lead where only Claude could have seen the
outcomes. Instead the lead is smallest there, and the window statistic is $-0.227$ [$-0.447$, $-0.042$]
(Figure~\ref{fig:post}b, Table~\ref{tab:windows}).

The two comparisons with Gemini point the same way, less precisely. On trials that became final in
April--June 2026, which Claude could have seen and Gemini could not, Claude leads Gemini by $0.074$,
against $0.302$ after July (difference $-0.228$ [$-0.468$, $+0.013$]). On trials that became final
between December 2025 and March 2026, which Gemini could have seen and GPT-5.5 could not, Gemini trails
GPT-5.5 by $0.122$, against $0.065$ after July (difference $-0.057$ [$-0.254$, $+0.154$]).

We read these results as evidence that remembered outcomes do not drive the differences between models,
not as proof that the models remember nothing. Every comparison uses the same 43 trials, only 9 of them
stopped, as its reference, and the large leads on those trials are probably partly noise. ``Could have
seen'' is not ``saw'': the last months before a cutoff are thinly represented in training data, and
Gemini's model card notes that some domains may end as early as January 2025. Finally, the tests concern
leads between models, not recall itself, which Section~\ref{sec:recog} establishes directly.

\subsection{Calibration and run-to-run variation}
\label{sec:calib}

\begin{figure}[t]
\centering
\includegraphics[width=\textwidth]{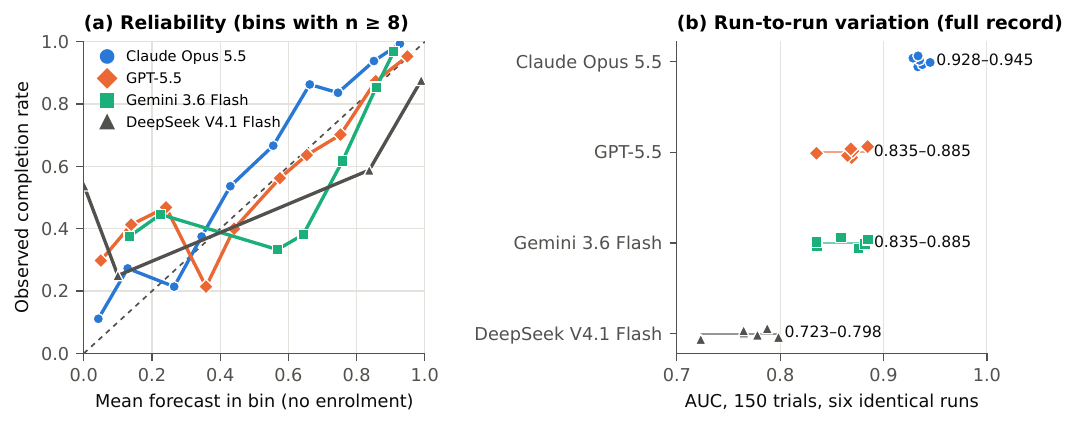}
\caption{(a) Reliability of forecasts without enrolment (10 equal-width bins; bins with at least 8
trials shown). (b) AUC of the full-record prompt on the same 150 trials across six identical runs.}
\label{fig:calib}
\end{figure}

Every LLM under-predicts completion: mean forecasts range from $0.658$ (Claude) to $0.714$ (GPT-5.5)
against an observed rate of $0.754$ (Table~\ref{tab:h2h}, Figure~\ref{fig:calib}a). Even uncalibrated,
Claude, GPT-5.5 and Gemini beat the base-rate Brier score ($0.101$--$0.130$ against $0.193$). DeepSeek,
which answers exactly 0 or 100 for 85\% of trials in its default thinking mode, does worse than the base
rate (Brier $0.223$) until Platt scaling repairs it to $0.148$. Calibrated Claude is the best forecaster
on Brier score ($0.091$).

Repeated runs move a trial's forecast by about 2 percentage points (median standard deviation 1.8--2.2
for every model). Across six runs, Claude's AUC on the 150-trial subset ranges from $0.928$ to $0.945$, and
GPT-5.5's and Gemini's from $0.835$ to $0.885$. DeepSeek behaves differently. Of its forecasts, 43\% never
move across the six runs, but 38\% of those that do move jump between 0 and 100 (21\% of the subset), and
its AUC ranges from $0.723$ to $0.798$ (Figure~\ref{fig:calib}b). Single-run comparisons involving DeepSeek
are therefore fragile.

\subsection{Label sensitivity}
\label{sec:labelsens}

\begin{table}[t]
\centering
\caption{Label sensitivity: full-record forecasts re-scored under each definition (Table~\ref{tab:labels}). The tree is retrained under each definition, and the constant is each definition's training completion rate. Platt rows give cross-fitted Brier scores.}
\label{tab:labelsens}
\small
\begin{tabular}{lcccccc}
\toprule
 & \multicolumn{2}{c}{A ($n=675$, rate 0.754)} & \multicolumn{2}{c}{B ($n=578$, rate 0.881)} & \multicolumn{2}{c}{C ($n=541$, rate 0.941)} \\
\cmidrule(lr){2-3}\cmidrule(lr){4-5}\cmidrule(lr){6-7}
Forecaster & AUC & Brier & AUC & Brier & AUC & Brier \\
\midrule
Constant (training rate) & 0.500 & 0.193 & 0.500 & 0.106 & 0.500 & 0.057 \\
Gradient-boosted tree (retrained) & 0.716 & 0.159 & 0.637 & 0.101 & 0.425 & 0.059 \\
\midrule
Claude Opus 5.5 & 0.923 & 0.101 & 0.904 & 0.099 & 0.931 & 0.089 \\
\quad Platt & & 0.091 & & 0.068 & & 0.041 \\
GPT-5.5 & 0.874 & 0.129 & 0.868 & 0.111 & 0.910 & 0.095 \\
\quad Platt & & 0.123 & & 0.081 & & 0.052 \\
Gemini 3.6 Flash & 0.868 & 0.130 & 0.858 & 0.112 & 0.886 & 0.096 \\
\quad Platt & & 0.125 & & 0.082 & & 0.049 \\
DeepSeek V4.1 Flash & 0.767 & 0.223 & 0.734 & 0.218 & 0.763 & 0.201 \\
\quad Platt & & 0.148 & & 0.095 & & 0.055 \\
\bottomrule
\end{tabular}
\end{table}

The ranking of the four models is the same under every label definition, and every LLM beats the
retrained tree on AUC (Table~\ref{tab:labelsens}). Brier scores are less stable. Under definition C,
where 94.1\% of test trials complete, no raw forecast beats the constant ($0.057$), but after Platt scaling
all four do ($0.041$--$0.055$). Under definition B, raw Claude beats the constant ($0.099$ against
$0.106$), and the other models do so only after calibration. A looser keyword rule for C, which counts
any ``lack of'' as efficacy, also catches 8 test-set stops for lack of accrual, funding or data; under
that rule Claude's AUC is $0.933$ and the conclusions are unchanged.

\subsection{Two independent collection runs}
\label{sec:revision}

The results above come from the second of two collections on the same trials. The first, on 8 October
2026, queried every model through a third-party API router. Three differences between the runs matter:
\begin{itemize}\itemsep2pt
  \item \textbf{Model access.} The earlier run reproduces three of the four models
  (Table~\ref{tab:replication}); the Gemini identifier is now served by Gemini 3.6 Flash.
  \item \textbf{A DeepSeek result that did not replicate.} In the earlier run, DeepSeek ranked the newest
  stratum better without enrolment than with it ($0.710$ against $0.686$); this run gives $0.682$ against
  $0.781$. Since DeepSeek's two runs correlate at only $r=0.60$--$0.67$ per trial, the earlier result was
  most likely noise.
  \item \textbf{Parsing and inputs.} The earlier harness took the first number in short replies and the
  last in long ones, and it misread some replies that began with a trial code or a bolded number; this
  run requires a fixed answer line. The earlier prompts also left the arm count empty for every trial,
  and the earlier harness stored only the first 80 characters of each reply; this run fills in the arm
  count and stores every reply in full.
\end{itemize}

\section{Discussion}

\paragraph{Audit the inputs against the registry's history.} Enrolment looks like design information but
is written after the outcome, and it supplies about three-quarters of a standard structured baseline's
skill above chance. Because the leak inflates the baseline more than the LLMs, a comparison that feeds
actual enrolment to both understates the LLMs' relative advantage while overstating the skill of every
forecaster. Registries keep their own histories, so the check costs only a download: compare each field
as first registered with its current value, and score each change against the label.

\paragraph{Assume the model knows the trial, and measure it.} Every model we tested names most trials
from their design alone and recalls many outcomes. A recognition probe that asks for the identifier is
cheap, and a correct eight-digit identifier is hard evidence where a model's yes or no is not. In-sample
AUC on registered trials should be reported as an upper bound.

\paragraph{Date the outcome, not the paper trail.} A holdout built on when results were posted measured
the leak, not memory, and the non-LLM control inherited the problem rather than solving it. Outcomes dated
after a model's cutoff provide the honest test, and comparisons between models with different cutoffs
provide a second one. Here all three such comparisons point away from memorisation, but the samples are
small and the post-cutoff trials differ in kind from the main sample.

\paragraph{What skill remains.} On trials resolved after GPT-5.5's cutoff, the four models score
$0.57$--$0.72$. Claude's $0.722$ is clearly above chance; its lead over a design-only tree on the same
trials is large but uncertain; and its lead over the other models does not look like recall. This is a
real but modest forecasting signal, well short of the in-sample $0.92$. Probabilities also need checking:
every model under-predicts completion, and one is worse than a constant until calibrated.

\section{Limitations}
\label{sec:limitations}

\begin{enumerate}\itemsep3pt
  \item \textbf{The label measures completion, not success.} Only 15.7\% of stops cite efficacy or
  futility, although the stricter definitions preserve the ranking of models.
  \item \textbf{The sample is selected.} Only trials that posted results enter the main sample.
  \item \textbf{The post-cutoff sets are small and different.} The July set has only 9 stopped trials,
  most of them stopped for business reasons. A registry status can change weeks or years after the real
  stop (27 of the 136 December-set trials reached primary completion before 2024), so some of these
  outcomes were knowable earlier; the stricter sets address this.
  \item \textbf{Cutoffs are bounds.} We use the training cutoffs the providers publish. DeepSeek does not
  state one, Gemini's model card notes that some domains end earlier, and data near any cutoff is sparse.
  \item \textbf{Recognition is a lower bound.} A model can know a trial without returning its identifier,
  and the identifier-and-title prompt explicitly invites recall.
  \item \textbf{Other leakage may remain.} We audited enrolment and compared every field across versions,
  but free text can still carry outcome information. In the test set, 91 eligibility texts and 15
  primary-outcome texts contain a word beginning with ``terminat'' or ``discontinu'', mostly referring to
  prior therapy or to stopping treatment after adverse events.
  \item \textbf{Settings are not fully controllable.} Claude's API does not accept a temperature setting
  and GPT-5.5 is a reasoning model, so both ran at default settings, and DeepSeek's thinking mode may
  ignore the temperature it is sent. The repeated runs measure the resulting noise rather than remove it.
  \item \textbf{Models change behind identifiers}, as Gemini's did between our two collection runs. We
  release every reply so that each number can be checked without re-querying.
  \item \textbf{Many comparisons.} Strata, variants and labels yield many comparisons, which we do not
  correct for multiplicity. We rest the conclusions on a few of them and report intervals throughout.
\end{enumerate}

\section{Conclusion}

How much of an LLM's apparent skill at forecasting Phase~3 cancer trial completion survives a careful
evaluation? Less than the in-sample numbers suggest, but more than nothing. The structured baseline draws
most of its skill from a field that records the outcome. The models recognise most trials and recall many
outcomes, so a holdout based on when results were posted cannot rule recall out. On outcomes dated after
the training cutoffs, the best model still ranks trials at an AUC of about 0.72, and the differences
between models are not where memorisation would put them. The general lesson extends beyond clinical
trials: check every input against its history, test whether the model knows the item, and measure skill
on outcomes the model could not have seen.

\section*{Reproducibility}
Every number, table and figure is computed by the released scripts from the released files, and no step
requires re-querying a model.
\begin{quote}\ttfamily\small
data/trials.jsonl \hfill 2{,}366 trials (registry snapshot, 8 Oct 2026)\\
data/ctgov\_history\_2026-10-11.json.gz \hfill every version of every record\\
data/prompts\_*.jsonl \hfill prompts rebuilt from the histories\\
data/postcutoff.jsonl \hfill 243 trials resolved after 1 Dec 2025\\
runs/ \hfill the earlier collection run (8 Oct 2026)\\
runs2/ \hfill every reply in this run, in full (28{,}272 calls)\\
rerun.py, config.json \hfill harness: prompts, parser, settings\\
rebuild/*.py \hfill prompts, post-cutoff sets and baselines from the histories\\
scripts/compute.py, tables.py, figs.py \hfill every statistic, table and figure
\end{quote}
\rmfamily\small
Bootstrap intervals use 2{,}000 resamples (4{,}000 for the window statistics) with fixed seeds. Registry
counts shift as the registry updates, so every count should be cited with its retrieval date.
\normalsize

\section*{Data and Code Availability}
Trial data are derived from the public ClinicalTrials.gov registry. Code, data, prompts and every model
reply are provided as supplementary material and will be deposited in a public archive on publication.

\section*{Ethics and Disclosure}
This study analyses public registry records about trials, not data about people. All claims concern the
behaviour of language and machine-learning models; nothing here is a claim about the efficacy or safety of
any treatment or about the conduct of any trial. An AI assistant, Claude (Anthropic), from one of the
evaluated model families, was used to write analysis code and to draft and edit the text; the author
reviewed the analyses and is responsible for all content. No competing interests are declared.

\begin{nonanon}
\section*{Acknowledgments}
The author thanks Colin F. Camerer and Pietro Perona for inspiring this project.
\end{nonanon}

\bibliography{references}
\bibliographystyle{tmlr}

\appendix
\section{Prompts}
\label{app:prompt}
The forecasting variants share a header and an answer line:
\begin{quote}\small\ttfamily
You are forecasting the outcome of a Phase 3 cancer clinical trial. Estimate the probability that this
trial will be COMPLETED (rather than terminated or withdrawn), as an integer from 0 to 100.\\[4pt]
Phase: \{phase\}\\ Condition: \{conditions\}\\ Interventions: \{type name; \ldots\}\\
Primary endpoints: \{measures\}\\ Enrolment: \{enrollment\}\\ Sponsor type: \{class\}\\
Arms: \{n\}\\ Allocation: \{allocation\}\\ Primary purpose: \{purpose\}\\[4pt]
Eligibility criteria (truncated to 6000 characters):\\ \{eligibility\}\\[4pt]
Reply with exactly one line and nothing else, in this form:\\
ANSWER: <integer from 0 to 100>
\end{quote}
\rmfamily\normalsize
Variant N drops the \texttt{Enrolment} line; A drops the eligibility block; B keeps only the eligibility
block; and X keeps only the condition and interventions. Variant I replaces the body with ``Consider the
registered clinical trial below. If you recall it, use what you remember about its outcome'', followed by
the NCT identifier and title. The recognition probe (R) reads:
\begin{quote}\small\ttfamily
Below is the design of a registered Phase 3 clinical trial, without its identifier or title. Do you
recognise this specific trial? Answer only from what you already know.\\[4pt]
\{the structured fields without enrolment, and the eligibility block\}\\[4pt]
Reply with one JSON object and nothing else, with these keys: \{"recognised": true or false,
"trial\_name": string or null, "nct\_id": string or null, "recalled\_outcome": "completed" or
"terminated" or "unknown", "confidence": integer 0-100\}
\end{quote}
\rmfamily\normalsize

\section{Additional tables}
\label{app:tables}
\begin{table}[!htbp]
\centering
\caption{Prompt ablation on the 675 test trials: AUC [95\% CI] by variant, and paired differences. DeepSeek's eligibility-only variant is scored on 672 trials (three replies hit the token cap).}
\label{tab:ablation}
\scriptsize
\setlength{\tabcolsep}{3pt}
\resizebox{\textwidth}{!}{%
\begin{tabular}{lcccc}
\toprule
Variant & Claude Opus 5.5 & GPT-5.5 & Gemini 3.6 Flash & DeepSeek V4.1 Flash \\
\midrule
Full record & 0.923 [0.900, 0.943] & 0.874 [0.843, 0.901] & 0.868 [0.837, 0.897] & 0.767 [0.728, 0.803] \\
No enrolment & 0.899 [0.872, 0.924] & 0.844 [0.812, 0.876] & 0.828 [0.796, 0.862] & 0.716 [0.672, 0.759] \\
Structured fields only & 0.921 [0.899, 0.941] & 0.859 [0.826, 0.889] & 0.843 [0.807, 0.876] & 0.770 [0.731, 0.808] \\
NCT number + title & 0.887 [0.859, 0.910] & 0.835 [0.805, 0.867] & 0.813 [0.775, 0.848] & 0.717 [0.673, 0.759] \\
Condition + interventions & 0.870 [0.839, 0.898] & 0.810 [0.770, 0.847] & 0.783 [0.744, 0.820] & 0.734 [0.691, 0.774] \\
Eligibility text only & 0.869 [0.836, 0.898] & 0.801 [0.762, 0.841] & 0.760 [0.719, 0.800] & 0.705 [0.662, 0.751] \\
\midrule
No enrolment $-$ NCT + title & +0.012 [$-$0.004, +0.029] & +0.008 [$-$0.017, +0.033] & +0.015 [$-$0.009, +0.041] & $-$0.001 [$-$0.047, +0.047] \\
No enrolment $-$ condition + interventions & +0.029 [+0.008, +0.050] & +0.034 [+0.008, +0.061] & +0.045 [+0.016, +0.074] & $-$0.017 [$-$0.057, +0.022] \\
\bottomrule
\end{tabular}}
\end{table}

\begin{table}[!htbp]
\centering
\caption{AUC by results-posting stratum, full record (C) and enrolment withheld (N), and the paired gain from enrolment (C $-$ N) in the oldest and newest strata.}
\label{tab:strata}
\scriptsize
\setlength{\tabcolsep}{3pt}
\begin{tabular}{lcccccccc}
\toprule
 & \multicolumn{2}{c}{$\le$2022 (295)} & \multicolumn{2}{c}{2023--24 (243)} & \multicolumn{2}{c}{$\ge$2025 (137)} & \multicolumn{2}{c}{Gain from enrolment} \\
\cmidrule(lr){2-3}\cmidrule(lr){4-5}\cmidrule(lr){6-7}\cmidrule(lr){8-9}
Forecaster & C & N & C & N & C & N & $\le$2022 & $\ge$2025 \\
\midrule
Gradient-boosted tree & 0.789 & 0.570 & 0.735 & 0.564 & 0.550 & 0.517 &  &  \\
Claude Opus 5.5 & 0.949 & 0.915 & 0.908 & 0.887 & 0.887 & 0.886 & +0.034 [+0.011, +0.062] & +0.002 [$-$0.034, +0.040] \\
GPT-5.5 & 0.920 & 0.855 & 0.842 & 0.834 & 0.816 & 0.839 & +0.065 [+0.025, +0.110] & $-$0.023 [$-$0.069, +0.025] \\
Gemini 3.6 Flash & 0.892 & 0.835 & 0.852 & 0.829 & 0.830 & 0.799 & +0.056 [+0.024, +0.089] & +0.032 [$-$0.029, +0.096] \\
DeepSeek V4.1 Flash & 0.767 & 0.741 & 0.755 & 0.707 & 0.781 & 0.682 & +0.027 [$-$0.049, +0.103] & +0.099 [+0.008, +0.187] \\
\bottomrule
\end{tabular}
\end{table}

\begin{table}[!htbp]
\centering
\caption{Left: an earlier, independent collection run (8 October 2026, through a third-party API router, with a looser answer parser) against this run, on the same trials (AUC; $r$ = per-trial correlation of forecasts). GPT-5.5's earlier no-enrolment run has 673 forecasts. Right: AUC on the current record minus AUC on the record as first registered, which carries planned rather than actual enrolment (paired 95\% CI).}
\label{tab:replication}
\scriptsize
\setlength{\tabcolsep}{3pt}
\begin{tabular}{lcccccccc}
\toprule
 & \multicolumn{3}{c}{Full record} & \multicolumn{3}{c}{Enrolment withheld} & \multicolumn{2}{c}{Current $-$ first registered} \\
\cmidrule(lr){2-4}\cmidrule(lr){5-7}\cmidrule(lr){8-9}
Model & Earlier & This run & $r$ & Earlier & This run & $r$ & Full & No enrolment \\
\midrule
Claude Opus 5.5 & 0.921 & 0.923 & 0.96 & 0.894 & 0.899 & 0.94 & +0.035 [+0.021, +0.050] & +0.009 [$-$0.001, +0.022] \\
GPT-5.5 & 0.868 & 0.874 & 0.89 & 0.850 & 0.844 & 0.89 & +0.035 [+0.012, +0.059] & +0.005 [$-$0.013, +0.024] \\
Gemini 3.6 Flash & 0.815 & 0.868 & 0.80 & 0.818 & 0.828 & 0.78 & +0.045 [+0.022, +0.069] & +0.010 [$-$0.010, +0.030] \\
DeepSeek V4.1 Flash & 0.749 & 0.767 & 0.67 & 0.705 & 0.716 & 0.60 & +0.041 [$-$0.003, +0.087] & $-$0.014 [$-$0.054, +0.025] \\
\bottomrule
\end{tabular}
\end{table}

\begin{table}[!htbp]
\centering
\caption{Window tests. Each row compares a model that could have seen trials resolved in the window (the first) with one that could not, using the published cutoffs in Table~\ref{tab:models}. Lead = AUC of the first minus AUC of the second, without enrolment. ``After'' = the 43 trials (9 stopped) whose status first became final from 1 July 2026, after every published cutoff. Memorisation predicts a positive difference. Paired 95\% intervals within each set; independent resamples (4{,}000) for the difference.}
\label{tab:windows}
\scriptsize
\setlength{\tabcolsep}{3pt}
\resizebox{\textwidth}{!}{%
\begin{tabular}{llcccc}
\toprule
Comparison & Window ($n$; stopped) & Lead, test set & Lead in window & Lead after & Window $-$ after \\
\midrule
Claude Opus 5.5 $-$ GPT-5.5 & 1 Dec 2025 -- 30 Jun 2026 (93; 26) & +0.055 [+0.032, +0.080] & +0.010 [$-$0.068, +0.092] & +0.237 [+0.064, +0.437] & $-$0.227 [$-$0.447, $-$0.042] \\
Claude Opus 5.5 $-$ Gemini 3.6 Flash & 1 Apr -- 30 Jun 2026 (43; 11) & +0.070 [+0.045, +0.096] & +0.074 [$-$0.070, +0.222] & +0.302 [+0.113, +0.508] & $-$0.228 [$-$0.468, +0.013] \\
Gemini 3.6 Flash $-$ GPT-5.5 & 1 Dec 2025 -- 31 Mar 2026 (50; 15) & $-$0.016 [$-$0.045, +0.011] & $-$0.122 [$-$0.264, +0.012] & $-$0.065 [$-$0.235, +0.081] & $-$0.057 [$-$0.254, +0.154] \\
\bottomrule
\end{tabular}}
\end{table}

\begin{table}[!htbp]
\centering
\caption{Stricter post-cutoff sets: cancer conditions, status first final after the date, and final primary completion no earlier than six months before it. As in Table~\ref{tab:post}, trials already in the 2{,}366-trial dataset are excluded (4 from the December set).}
\label{tab:strict}
\scriptsize
\setlength{\tabcolsep}{3pt}
\begin{tabular}{lccccc}
\toprule
Set & $n$ (stopped) & Claude Opus 5.5 & GPT-5.5 & Gemini 3.6 Flash & DeepSeek V4.1 Flash \\
\midrule
From 1 Dec 2025, strict & 83 (27) & 0.737 [0.631, 0.833] & 0.667 [0.540, 0.778] & 0.608 [0.486, 0.725] & 0.590 [0.471, 0.709] \\
From 1 Jul 2026, strict & 28 (7) & 0.833 [0.667, 0.963] & 0.599 [0.331, 0.841] & 0.514 [0.292, 0.743] & 0.537 [0.307, 0.759] \\
\bottomrule
\end{tabular}
\end{table}

\begin{table}[!htbp]
\centering
\caption{Baselines on the test trials. Every row except the first withholds actual enrolment. Text model: word and character $n$-gram TF-IDF with logistic regression. Cross-fitting by drug programme keeps all trials of a programme on one side of each fold. The rolling origin trains on trials that started by 2018 and tests on the 286 test trials that started in 2019 or later.}
\label{tab:ladder}
\scriptsize
\begin{tabular}{lrc}
\toprule
Baseline & $n$ & AUC [95\% CI] \\
\midrule
Tree, actual enrolment & 675 & 0.716 [0.666, 0.763] \\
Tree, enrolment withheld & 675 & 0.556 [0.504, 0.604] \\
Tree, planned enrolment and arm count as first registered & 675 & 0.566 [0.517, 0.614] \\
Text model (TF-IDF), trained on starts $\le$2015 & 675 & 0.580 [0.531, 0.630] \\
Text model, cross-fitted on test trials, folds by drug programme & 675 & 0.597 [0.549, 0.642] \\
Text model, cross-fitted on test trials, random folds & 675 & 0.667 [0.622, 0.712] \\
\midrule
Rolling origin: tree, trained on starts $\le$2018 & 286 & 0.606 [0.537, 0.674] \\
Rolling origin: tree with planned enrolment & 286 & 0.613 [0.542, 0.678] \\
Rolling origin: text model & 286 & 0.616 [0.537, 0.683] \\
Same 286 trials: Claude Opus 5.5, no enrolment & 286 & 0.883 [0.838, 0.920] \\
Same 286 trials: GPT-5.5, no enrolment & 286 & 0.836 [0.786, 0.882] \\
Same 286 trials: Gemini 3.6 Flash, no enrolment & 286 & 0.794 [0.740, 0.847] \\
Same 286 trials: DeepSeek V4.1 Flash, no enrolment & 286 & 0.688 [0.627, 0.748] \\
\bottomrule
\end{tabular}
\end{table}

\end{document}